\documentclass[aps,pra,superscriptaddress,showpacs,twocolumn]{revtex4-2}
\usepackage{bm}
\usepackage{amsmath}
\usepackage{amssymb}
\usepackage{slashed}
\usepackage{float}
\allowdisplaybreaks
\usepackage{subcaption}
\usepackage{dcolumn}
\usepackage{graphicx}

\usepackage{physics}
\usepackage{nicefrac}
\newcolumntype{w}[1]{D{.}{.}{#1}}
\newcommand{\Za}{Z\alpha}
\newcommand{\vare}{\varepsilon}
\newcommand{\balpha}{\vec{\alpha}}

\newcommand{\lbr}{\langle}
\newcommand{\rbr}{\rangle}

\newcommand{\bfr}{{\vec{r}}}

\newcommand{\hr}{{\hat{\vec{r}}}}

\begin{document}

\title{Nuclear magnetic shielding in helium-like ions}

\author{Vladimir A. Yerokhin}
\affiliation{Max-Planck Institute for Nuclear Physics,
Saupfercheckweg 1, 69117 Heidelberg, Germany}

\author{Krzysztof Pachucki}
\affiliation{Faculty of Physics, University of Warsaw,
             Pasteura 5, 02-093 Warsaw, Poland}

\author{Zolt\'an Harman}
\affiliation{Max-Planck Institute for Nuclear Physics,
Saupfercheckweg 1, 69117 Heidelberg, Germany}

\author{Christoph H. Keitel}
\affiliation{Max-Planck Institute for Nuclear Physics,
Saupfercheckweg 1, 69117 Heidelberg, Germany}

\begin{abstract}

{\em Ab initio} QED calculations of the nuclear magnetic shielding constant in helium-like ions
are presented.
We combine the nonrelativistic QED approach based on an expansion in
powers of the fine-structure constant $\alpha$ and the so-called ``all-order'' QED
approach which includes all orders in the parameter $\Za$ but uses a perturbation
expansion in the parameter $1/Z$ (where $Z$ is the nuclear charge number).
The combination of the two complementary methods
makes our treatment applicable both to low-$Z$ and high-$Z$ ions.
Our calculations confirm the presence of
a rare anti-screening effect for the relativistic shielding correction and demonstrate
the importance of the inclusion of the negative-energy part of the Dirac spectrum.

\end{abstract}

\maketitle

\section{Introduction}

Magnetic moments of nuclei are often determined from nuclear magnetic resonance (NMR) measurements.
Despite high precision of these experiments,
the accuracy of the extracted nuclear moments
is severely limited by the restricted knowledge of the magnetic shielding caused by the chemical surrounding.
Such effects are difficult to calculate reliably, which often led to significant deficiencies in the
literature data on nuclear magnetic moments \cite{raghavan:89,stone:05}. As an example,
the so-called “bismuth hyperfine puzzle”
\cite{ullmann:17,karr:17} was recently resolved \cite{skripnikov:18} and traced back to
an inaccuracy of the nuclear magnetic moment caused by
shortcomings in calculations of the shielding correction.

Much more accurate determinations of nuclear magnetic moments can nowadays be achieved by
Penning-trap measurements of
the combined Zeeman and hyperfine structure of few-electron atoms or ions. The shielding constants of
such systems can be calculated {\em ab initio}
within the framework of QED, with a detailed analysis of uncertainties
due to omitted higher-order effects. Precise determinations
of magnetic moments of a number of light nuclei
by this method were reported in the last years
\cite{puchalski:22,schneider:22,pachucki:23:limag,pachucki:10}.
In particular, the magnetic moment of the proton was accurately measured
by the Penning-trap technique in Ref.~\cite{mooser:14}. This technique can
in principle be extended to measurements of other nuclei and closed-shell ions.

Highly sophisticated calculations of the nuclear shielding
have been recently accomplished for the helium atom
\cite{rudzinski:09,wehrli:21,pachucki:23:shield}, motivated by perspectives of using
the hyperpolarized helium NMR probes as a new standard for absolute
magnetometry \cite{nikiel:14,fan:19,farooq:20}.
The calculation of Ref.~\cite{rudzinski:09} revealed a rare effect of {\em `anti-screening'}
for the relativistic shielding correction, corresponding
to the situation when
an effect for two correlated $(1s)^2$ electrons is larger than for two non-interacting $1s$ electrons.
This defies the physically intuitive picture in which each of the correlated electrons should experience
a slightly smaller nuclear charge because it is effectively screened by the second electron,
the effect commonly known as {\em screening}.
So far the presence of the anti-screening effect
has not been confirmed by an independent calculation.

The goal of the present study is to perform {\em ab initio} QED calculations of the nuclear
magnetic shielding of helium-like ions for a wide range of nuclear charges $Z$.
This will be achieved by merging together two complementary methods, namely,
the nonrelativistic QED (NRQED) approach based on an expansion in
powers of the fine-structure constant $\alpha$ and the so-called ``all-order'' QED
approach which includes all orders in the parameter $\Za$ but uses a perturbation
expansion in the parameter $1/Z$. The NRQED method alone is applicable
only to low-$Z$ ions, since the uncalculated higher-order effects scale with high powers
of $Z$. By contrast, the all-order method is effective in the high-$Z$ region, since
the $1/Z$ expansion converges fast there. In this work we unify
these two methods, so that the resulting approach becomes applicable for the whole range
of $Z$. For the first time such unified approach
was applied by G.~Drake for calculating energies and transition rates of
helium-like ions in Refs.~\cite{drake:79,drake:88:cjp}.

The outline of our calculations is as follows. First, we employ the
NRQED approach to calculate the leading shielding contribution of order $\alpha^2$
as well as the relativistic, nuclear and QED corrections of order $\alpha^4$, $\alpha^2m/M$,
and $\alpha^5\ln\alpha$. Then we address the higher-order
corrections within the all-order
method. We calculate the one-electron shielding contribution, the one-photon exchange,
QED, and the nuclear magnetization
distribution effects. By analysing the $\Za$ expansion of the individual corrections,
we identify the lowest-order contributions already included into the NRQED treatment and
remove them, thus avoiding double counting and obtaining
the final results for the shielding constant.

%The structure of the paper is as follows. Sec.~\ref{sec:1} describes calculations
%performed within the NREQD approach. In Sec.~\ref{sec:2} we compute
%the higher-order effects within the all-order method.
%Finally, in Sec.~\ref{sec:3} we combine all shielding contributions and obtain
%the final results.

\section{NRQED approach}
\label{sec:1}

Within the NRQED approach, the shielding constant $\sigma$ of low-$Z$ atoms is represented as
a double expansion in $\alpha$ and the electron-to-nucleus mass ratio $m/M$,
\begin{align}
\sigma = \alpha^2 \sigma^{(2)} + \alpha^4 \sigma^{(4)} + \alpha^2 \frac{m}{M} \sigma^{(2,1)}
 + \alpha^5 \sigma^{(5)} + \ldots\,.
\end{align}
As is customary in NRQED calculations, we will use the atomic units in the following formulas
in this Section.
To the leading order in $\alpha$ and zeroth order in $m/M$,
the shielding constant $\sigma$ takes the form \cite{ramsey:50:dia}
\begin{align}\label{2}
\sigma^{(2)} = \frac13\sum_{a} \Big< \frac1{r_a}\Big>\,,
\end{align}
where $r_a$ is the distance between the nucleus and $a$'th electron and the summation runs over all electrons.
The relativistic shielding correction
of order $\alpha^4$ was derived
in Ref.~\cite{rudzinski:09}, with the result
\begin{widetext}
\begin{align}\label{eq:s4}
\sigma^{(4)} = &\, \sum_a
\biggl\langle
\frac{1}{12\, r_a^3}\,
\Big(\frac{\vec r\cdot\vec r_1\;\vec r\cdot\vec r_2}{r^3} - 3\,\frac{\vec r_1\cdot\vec r_2}{r}\Big)
-\frac{1}{6}\,
 \Big( \frac{1}{r_a}\,p_a^2+\frac{(\vec r_a\times\vec p_a)^2}{r_a^3}+4\,\pi\,\delta(\vec r_a)
 \Big)
\biggl\rangle
 \nonumber \\ &
+ \frac{2}{3}\,\biggl\langle\biggl(\frac{1}{r_1}+\frac{1}{r_2}\biggr)\,
\frac{1}{(E-H)'}\,\biggl[\sum_a \biggl(\frac{\pi\,Z}{2}\,\delta(\vec r_a)-\frac{p_a^4}{8}\biggr)
+\pi\,\delta(\vec r)-\frac{1}{2}\,p_1^i\biggl(\frac{\delta^{ij}}{r}+\frac{r^i\,r^j}{r^2}\biggr)\,p_2^j
\biggr]\biggr\rangle
 \nonumber \\ &
-\frac{2}{9}\,\biggl\langle\pi\,\bigl[\delta(\vec
  r_1)-\delta(\vec r_2)\bigr]\,
\frac{1}{(E-H)}\,\biggl[
3\,p_1^2-3\,p_2^2-\frac{Z}{r_1}+\frac{Z}{r_2}-\frac{\vec r\cdot(\vec r_1+\vec r_2)}{r^3}
\biggr]\biggr\rangle
 \nonumber \\ &
-\frac{1}{6}\,\biggl\langle\biggl(
\frac{\vec r_1\times\vec p_1}{r_1^3} + \frac{\vec r_2\times\vec p_2}{r_2^3}\biggr)\,
\frac{1}{(E-H)}\biggl[
\vec r_1\times\vec p_1\,p_1^2 + \vec r_2\times\vec p_2\,p_2^2
+\frac{1}{r}\,\vec r_1\times\vec p_2 + \frac{1}{r}\,\vec r_2\times\vec p_1
-\vec r_1\times\vec r_2\,\frac{\vec r}{r^3}\cdot(\vec p_1+\vec p_2)\biggr]\biggr\rangle
 \nonumber \\ &
-\frac{1}{8}\,\biggl\langle
\biggl(\frac{r_1^i\,r_1^j}{r_1^5}-\frac{r_2^i\,r_2^j}{r_2^5}\biggr)^{(2)}\,
\frac{1}{(E-H)}\,\biggl(Z\,\frac{r_1^i\,r_1^j}{r_1^3} - Z\,\frac{r_2^i\,r_2^j}{r_2^3}
+\frac{r^i}{r^3}\,(r_1^j+r_2^j)\biggr)^{(2)}\biggr\rangle
\,,
\end{align}
\end{widetext}
where $\vec r\equiv \vec r_1-\vec r_2$, $(p^i\,q^i)^{(2)} =
p^i\,q^j/2+p^j\,q^i/2-\delta^{ij}\,\vec p\cdot\vec q/3$,
and $1/(E-H)'$ is the reduced Green function (with the reference state removed from the
sum over the spectrum).

The leading-order nuclear recoil correction was derived in Ref.~\cite{pachucki:08:recoil} and
later corrected in Ref.~\cite{pachucki:23:shield}. The result is
\begin{align}
\sigma^{(2,1)} \equiv &\, \frac{1-g_N}{g_N}\,\sigma^{(2,1)}_a + \sigma^{(2,1)}_b
 \nonumber \\ = &\,
 \frac{1-g_N}{g_N}\,\frac{ \lbr p_N^2 \rbr }{3Z}\,
 + \frac13 \Big< \sum_a \frac1{r_a} \frac1{(E-H)'} p_N^2\Big>
 \nonumber \\ &
 + \frac13 \Big< \sum_a \vec{r}_a \times \vec{p}_N \frac1{(E-H)}
                 \sum_b \frac{\vec{r}_b}{r_b^3} \times \vec{p}_b\Big>\,,
\end{align}
where $\vec{p}_N = -\sum_a\vec{p}_a$, $g_N = (M/m_p)(\mu/\mu_N)/(ZI)$
is the nuclear $g$-factor;
$M$, $I$ and $\mu$ are the nuclear mass, spin and the magnetic moment,
respectively; $m_p$ is the proton mass and $\mu_N$ is the nuclear magneton.

The logarithmic part of the leading QED correction of order $\alpha^5$
was derived in Ref.~\cite{rudzinski:09}. We write it as
\begin{align}
\sigma^{(5)}_{\rm log} = &\,
 \ln (\Za)\,\bigg[
   -\frac{16 Z}{9}\, \Big< \sum_a\frac1{r_a}\, \frac1{(E-H)'}\, \sum_b\delta(\vec{r}_b)\Big>
\nonumber \\ &
   +\frac{28}{9}\, \Big< \sum_a\frac1{r_a}\, \frac1{(E-H)'}\, \delta(\vec{r})\Big>
   - \frac{40}{9}\, \Big< \sum_a\delta(\vec{r}_a)\Big>
   \bigg]\,.
\end{align}
This formula differs from the one from Ref.~\cite{rudzinski:09} by $\ln Z$ in the
second-order correction
containing $\delta(\vec{r})$. In obtaining it, we took into account that
the two-electron Lamb shift contains, in addition to $\ln \alpha$, an implicit
$\ln Z$ term usually hidden in the Araki-Sucher term $\sim \lbr r^{-3}\rbr$
\cite{drake:85}.
The nonlogarithmic QED contribution of order $\alpha^5$ was derived and calculated for helium
in Ref.~\cite{wehrli:21}. Its numerical calculation is rather complicated as it involves
perturbations of the so-called Bethe logarithm. For this reason, we will address this and
higher-order QED corrections within the $1/Z$ expansion in the next Section.

Numerical calculations of the NRQED corrections summarized above
were carried out with the basis set of exponential functions
$e^{-\alpha_i\,r_1-\beta_i\,r_2-\gamma_i\,r}$ \cite{korobov:00}.
The method of calculations was developed in our previous investigations,
see Ref.~\cite{yerokhin:21:hereview} for a review.
The most computationally intense part was the calculation of $\sigma^{(4)}$.
While the evaluation of first-order matrix elements in Eq.~(\ref{eq:s4}) was
relatively straightforward, the computation of second-order matrix elements
turned out to be rather demanding. To achieve a high numerical accuracy,
we used carefully optimized basis sets for the intermediate electron
states. The optimization was carried out for symmetric second-order corrections,
using non-uniform distributions of nonlinear basis-set parameters (see 
Ref.~\cite{yerokhin:21:hereview} for details), with a typical size
of the basis set $N = 1200$. The obtained basis sets were 
then used for computation of non-symmetric second-order corrections in 
Eq.~(\ref{eq:s4}).

Results of our numerical calculations of $\sigma^{(2)}$, $\sigma^{(2,1)}$, $\sigma^{(4)}$,
and $\sigma^{(5)}_{\rm log}$
for $Z\le 12$ are presented in Table~\ref{tab:NRQED}.
The corresponding values for $Z>12$ can be readily obtained
by using the $1/Z$ expansion, which is of the form
\begin{align}
\frac{\sigma^{(2)}}{Z} = \sum_{k=0}^{\infty} \frac{c_k}{Z^k}\,,
\end{align}
and similarly for other corrections. The leading coefficients $c_0$
are known analytically from the hydrogen theory,
whereas the higher-order coefficients $c_k$ were obtained by
fitting our numerical results.

For the relativistic correction $\sigma^{(4)}$ we find a small deviation
from the helium result of Ref.~\cite{rudzinski:09}. The difference comes from
the second-order contribution with the $^3D$ intermediate states,
labeled as $Q_{12}$ in Ref.~\cite{rudzinski:09}. The deviation
is small and does not influence the final theoretical prediction for the
helium shielding constant within its estimated error.

Our results summarized in Table~\ref{tab:NRQED} confirm the previous
findings \cite{rudzinski:09} of the presence of the unusual
anti-screening effect for $\sigma^{(4)}$. Indeed, the absolute values
of $\sigma^{(4)}/Z^3$ for all nuclear charges are larger than
the corresponding limiting value for noninteracting electrons,
$c_0 = \nicefrac{97}{54} \approx 1.796$. This is in contrast to
all other corrections examined in Table~\ref{tab:NRQED}, which
exhibit the normal screening effect.
It is important to note that the first two $1/Z$-expansion coefficients $c_0$ and
$c_1$ of  $\sigma^{(4)}$ are independently cross-checked by our calculation of the one-photon
exchange correction in Sec.~\ref{sec:2}, thus excluding a possibility of a
technical mistake in the derivation of $\sigma^{(4)}$.
The probable explanation of the anti-screening effect is the singlet-triplet
mixing. Specifically, the relativistic effects mix the ground $1^1S_0$
state with intermediate $n^3S$ states. This mixing is quite large and
changes the behaviour of the relativistic correction in the case of helium-like
atoms as compared to the hydrogen-like case.

\begin{table*}
\caption{NRQED shielding corrections for different nuclear charges $Z$ and
their $1/Z$-expansion coefficients $c_k$.
\label{tab:NRQED}}
\begin{ruledtabular}
\begin{tabular}{cw{3.13}w{3.10}w{3.10}w{3.10}w{3.10}}
   $Z$ & \multicolumn{1}{c}{$\sigma^{(2)}/Z$}
                & \multicolumn{1}{c}{$\sigma^{(4)}/Z^3$}
                    & \multicolumn{1}{c}{$\sigma^{(5)}_{\rm log}/[Z^3 \ln(\Za)]$}
    & \multicolumn{1}{c}{$\sigma^{(2,1)}_a/Z$}
                & \multicolumn{1}{c}{$\sigma^{(2,1)}_b/Z$}
   \\
\hline\\[-5pt]
  2  &  0.562\,772\,266\,9  & 2.321\,754\,4  & -0.710\,693\,3 &  0.510\,465\,64  & -0.597\,289\,84 \\
     &  0.562\,772\,266\,8^a& 2.321\,42^a    &  \\
  3  &  0.597\,316\,533     & 2.070\,397\,4  & -0.837\,418\,1 &  0.560\,658\,46  & -0.619\,929\,38 \\
  4  &  0.614\,625\,068     & 1.979\,393\,5  & -0.906\,430\,9 &  0.586\,503\,61  & -0.631\,434\,60 \\
  5  &  0.625\,021\,856     & 1.933\,018\,3  & -0.949\,438\,3 &  0.602\,232\,65  & -0.638\,394\,84 \\
  6  &  0.631\,957\,258     & 1.905\,101\,6  & -0.978\,729\,3 &  0.612\,807\,06  & -0.643\,058\,72 \\
  7  &  0.636\,912\,900     & 1.886\,523\,9  & -0.999\,940\,0 &  0.620\,402\,23  & -0.646\,401\,72 \\
  8  &  0.640\,630\,484     & 1.873\,300\,7  & -1.016\,000\,2 &  0.626\,121\,09  & -0.648\,915\,36 \\
  9  &  0.643\,522\,387     & 1.863\,423\,1  & -1.028\,579\,1 &  0.630\,582\,19  & -0.650\,874\,23 \\
 10  &  0.645\,836\,163     & 1.855\,771\,1  & -1.038\,696\,3 &  0.634\,159\,23  & -0.652\,443\,72 \\
 11  &  0.647\,729\,404     & 1.849\,672\,8  & -1.047\,009\,0 &  0.637\,091\,24  & -0.653\,729\,45 \\
 12  &  0.649\,307\,201     & 1.844\,700\,8  & -1.053\,960\,0 &  0.639\,538\,22  & -0.654\,801\,99 \\
 \hline
 $c_0$ &  \multicolumn{1}{c}{$\nicefrac{2}{3}$}    & \multicolumn{1}{c}{$\nicefrac{97}{54}$} &
                                                            \multicolumn{1}{c}{$-\nicefrac{32}{9\pi}$}
       &  \multicolumn{1}{c}{$\nicefrac{2}{3}$}    &  \multicolumn{1}{c}{$-\nicefrac{2}{3}$} \\
 $c_1$ &  \multicolumn{1}{c}{$-\nicefrac{5}{24}$}  & 0.514\,442\,6     &  0.947\,740 &  -0.327\,804 &  0.143\,12 \\
 $c_2$ &   0.000\,000\,0                           & 0.770\,0          & -0.159\,8   &   0.026\,44  & -0.008\,9 \\
 $c_3$ &   0.002\,899\,7                           & 0.288             & -0.104      &   0.008\,6   &  0.000\,3 \\
 $c_4$ &  -0.000\,592                              & 0.40              &  0.003 \\
 $c_5$ &  -0.001\,04                               &     \\
    \end{tabular}
\end{ruledtabular}
$^a$ Ref.~\cite{rudzinski:09}
\end{table*}

\section{All-order approach}
\label{sec:2}

In order to access the higher-order effects $\sim\!\alpha^{5+}$,
we will adopt the so-called all-order QED approach. This method includes all orders in
the parameter $\Za$ but expands
in the electron-electron interaction, with the expansion
parameter $1/Z$. In order to separate out the higher-order contributions
beyond what is already included into the NRQED treatment in Sec.~\ref{sec:1},
we will examine the $\Za$ expansion
of the all-order results and remove the double counting by subtracting
the leading-order contributions.
The zeroth order in $1/Z$ is delivered by the independent-particle
approximation, which neglects the interaction between the electrons.
Further terms of the $1/Z$ expansion are described by Feynman diagrams containing
an exchange by one, two, etc. virtual photons between the electrons.
In this Section we will use the relativistic units ($\hbar = c = 1$).

\subsection{Electron-structure effects}
We start with examining the so-called electron-structure effects, which
are induced by Feynman diagrams without radiative loops.
\subsubsection{One-electron}
In the independent-particle approximation, the relativistic shielding constant
for the $(1s)^2$ state of the helium-like ion is,
see Ref.~\cite{yerokhin:12:shield} for details,
\begin{align}     \label{eq:6}
\sigma_{\rm rel, 1el} = \alpha  \sum_{\mu_a}
\sum_{n \ne a}
 \frac1{\vare_a-\vare_n}\,
  \lbr a| {V}_{g} |n\rbr
  \lbr n| {V}_{h} |a\rbr\,,
\end{align}
where $\mu_a$ is the momentum projection of the $1s$ electron,
the sum over $n$ runs over the complete spectrum of the Dirac equation, and
$V_g$ and $V_h$ are effective interactions of a Dirac electron
with the external magnetic field ($V_g$) and
with the magnetic dipole nuclear field ($V_h$),
\begin{align}
{V}_{g} = (\bfr\times\balpha)_z\,, \  \ \mbox{\rm and}\ \
{V}_{h} = \frac{\displaystyle
  (\bfr\times\balpha)_z}{\displaystyle r^3}\,.
\end{align}

For the point nuclear charge, Eq.~(\ref{eq:6}) can be calculated analytically
\cite{moore:99,pyper:99:a,pyper:99:b,moskovkin:04,ivanov:09}, with the result
\begin{eqnarray} \label{eq5a}
\sigma_{\rm rel, 1el} &=& 2\Big(-\frac{4\alpha\Za}{9}\Big) \left(\frac13-\frac1{6(1+\gamma)}
 +\frac{2}{\gamma}-\frac3{2\gamma-1} \right)
  \nonumber \\
  &=& 2\alpha\Za \bigg[\frac13 + \frac{97}{108}(\Za)^2
   + \frac{289}{216}(\Za)^4
  \nonumber \\ &&
    + \frac{3269}{1728}(\Za)^6
     + \ldots   \bigg]\,,
\end{eqnarray}
where $\gamma = \sqrt{1-(\Za)^2}$.
For an extended nuclear charge distribution, Eq.~(\ref{eq:6}) can be readily
calculated numerically with help of the dual-kinetic balance finite
basis set method \cite{shabaev:04:DKB}.

The higher-order one-electron relativistic correction is obtained from the above expression
by subtracting
the first two terms of the $\Za$ expansion,
\begin{align}
\sigma_{\rm rel, 1el}^{\rm h.o.} = &\, \sigma_{\rm rel, 1el} - 2\alpha(\Za)\Big[ \frac13 + \frac{97}{108}(\Za)^2\Big]
%  \\ &
%  = 2\alpha(\Za)^5\,\frac{289}{216} + \ldots
  \,.
\end{align}

It should be noted that when performing the summation over the Dirac energy spectrum in
Eq.~(\ref{eq:6}), the inclusion of the negative-energy part of the spectrum is mandatory,
as its contribution is very large, especially for low-$Z$ ions.
This is explained by the fact that the nonrelativistic limit of
the nuclear shielding constant in atoms is induced solely by
the negative-energy part of the Dirac spectrum. So,
for $Z = 2$, the negative-energy states induce 99.9\% of the total result.
With the increase of $Z$, the relative contribution of the negative-energy states
gradually diminishes but is still very significant, e.g., for $Z = 60$ it is 37\%.
This is in sharp contrast to calculations of transition
energies, where the negative-energy contribution is suppressed by a factor of $(\Za)^3$
compared to the leading nonrelativistic result \cite{sucher:80}.

\subsubsection{One-photon exchange}

The one-photon exchange correction to the nuclear magnetic shielding can be
obtained as a perturbation of
the one-photon exchange correction to the energy
by two external interactions, $V_g$ and $V_h$.
For the ground state of a helium-like ion, we obtain
\begin{widetext}
\begin{align}     \label{eq:9}
\sigma_{\rm rel, 1ph} = \alpha \sum_{P}(-1)^P\, &\  \biggl[
 \lbr PaPb|I|\delta_{hg}a\,b\rbr
+ \lbr PaPb|I|\delta_{gh}a\,b\rbr
  \nonumber \\
&+ \lbr PaPb|I|\delta_{h}a\,\delta_{g}b\rbr
+ \lbr \delta_{h}PaPb|I|\delta_{g}a\,b\rbr
+ \lbr \delta_{h}PaPb|I|a\,\delta_{g}b\rbr
  \nonumber \\
&- \lbr PaPb|I|ab\rbr \lbr\, \delta_ha|\delta_ga\rbr
- \lbr a|V_{g}|a\rbr\,\lbr PaPb|I|\widetilde{\delta}_ha\,b\rbr
- \lbr a|V_{h}|a\rbr\,\lbr PaPb|I|\widetilde{\delta}_ga\,b\rbr
\biggr]\,,
\end{align}
\end{widetext}
where $a$ and $b$ denote the two electrons in the $(1s)^2$ shell,
$P$ is the permutation operator ($PaPb = ab$ or $ba$), $(-1)^P$
is the sign of the permutation,
the perturbations of the wave functions
are defined by
\begin{align}\label{12}
|\delta_i a\rbr = \sum_{n}^{\vare_n\neq \vare_a} \frac{|n\rbr \lbr n|V_i|a\rbr}{\vare_a-\vare_n}\,,
\end{align}
\begin{align}
|\widetilde{\delta}_i a\rbr = \sum_{n}^{\vare_n\neq \vare_a} \frac{|n\rbr \lbr n|V_i|a\rbr}{(\vare_a-\vare_n)^2}\,,
\end{align}
\begin{align}\label{14}
|\delta_{hg} a\rbr = \sum_{n_1}^{\vare_{n_1}\neq \vare_a}\sum_{n_2}^{\vare_{n_2}\neq \vare_a}
 \frac{|n_1\rbr \lbr n_1|V_h|n_2\rbr\lbr n_2|V_g|a\rbr}{(\vare_a-\vare_{n_1})(\vare_a-\vare_{n_2})}\,,
\end{align}
and $|\delta_{gh} a\rbr$ is obtained from the last equation by interchanging
$g$ and $h$.
Furthermore, $I \equiv I(r_{12})$ denotes the electron-electron interaction operator
in the Coulomb gauge for the zero transferred energy,
\begin{align}
I(r_{12}) &\ = \frac{\alpha}{r_{12}}
 -\frac{\alpha}{2\,r_{12}}\,
    \left[ \balpha_1\cdot\balpha_2 + \left( \balpha_1\cdot \hr_{12}\right)
             \left( \balpha_2\cdot \hr_{12}\right) \right]\,,
\end{align}
where
$\balpha$ are the Dirac matrices, $\bfr_{12} =
\bfr_1-\bfr_2$,  and $\hr = \bfr/r$.
We note that in obtaining Eq.~(\ref{eq:9}) we took into account that the two electrons in the
$(1s)^2$ shell have the same energy, so that the frequency dependence of the
electron-electron interaction operator does not play any role in this case.

We calculated Eq.~(\ref{eq:9})
numerically with help of the $B$-spline
basis set method \cite{johnson:88} for the point-charge nuclear model
and with the dual-kinetic-balance method \cite{shabaev:04:DKB}
for the extended-charge nuclear model. The typical basis size 
used in the computation was $N = 100$ - $150$. 
In order to achieve a high numerical accuracy in the low-$Z$ region
(required for a high-precision fitting of the $d_2$ coefficient), we had to
use the quadruple-precision (appr. 32 decimal digits) in our computation.

The results for the point nuclear model are listed in Table~\ref{tab:1ph}.
In order to remove the double counting with the NRQED results,
we analyse the $\Za$-expansion of the one-photon
exchange correction, which is of the form
\begin{align}
\sigma_{\rm rel, 1ph} = \alpha^2 \sum_{k=0}^{\infty} d_{2k}\,(\Za)^{2k}\,.
\end{align}
The coefficients $d_{2k}$ obtained by fitting our numerical results are
listed in Table~\ref{tab:1ph}.
The first coefficient $d_0 = -\nicefrac{5}{24}$
corresponds to the $1/Z^1$ coefficient of expansion of $\sigma^{(2)}/Z$,
see Table~\ref{tab:NRQED}. The second coefficient $d_2$ corresponds to the
$1/Z^1$ coefficient of $\sigma^{(4)}/Z^3$, see Table~\ref{tab:NRQED}.
Other coefficients
represent contributions of order $\alpha^6$ and higher which have
not been accounted for in Sec.~\ref{sec:1}.

The higher-order one-photon exchange contribution is obtained as
\begin{align}
\sigma_{\rm rel, 1ph}^{\rm h.o.} = &\, \sigma_{\rm rel, 1ph} - \alpha^2\Big[ d_0 + d_2\,(\Za)^2\Big]
  \,,
\end{align}
where $d_0$ and $d_2$ are listed in Table~\ref{tab:1ph}.

Analysing contributions induced by the positive- and negative-energy parts
of the Dirac spectrum in the one-photon exchange shielding correction,
we observe that similarly to the one-electron case, the contribution
of the negative-energy states is very significant.
For $Z = 2$, they contribute 99.9\% of the
total result, whereas for $Z = 60$, the negative- and positive-energy
contributions are of the same magnitude and of the opposite sign.
This demonstrates the importance
of the proper treatment of the negative-energy part of the Dirac spectrum
in calculations of the nuclear shielding.
We note that a similar dominance of the negative-energy contribution
was recently found for the $M1$ polarizability in strontium
\cite{wu:23}.

\subsubsection{$\ge 2$ photon exchange}

The uncertainty due to two and more photon exchange is estimated basing
on the pattern of the
available $1/Z$-expansion coefficients of the $\sigma^{(4)}$ correction.
Specifically,
\begin{align}
\sigma_{\rm rel, 2ph+}^{\rm h.o.} \approx \pm \sigma_{\rm rel, 1ph}^{\rm h.o.}\, 
 2\,\frac{c_2}{c_1\,Z} 
 \approx \pm \sigma_{\rm rel, 1ph}^{\rm h.o.}\, \frac{3}{Z} \,,
\end{align}
where $2$ is the conservative factor.

\begin{table}
\caption{One-photon exchange shielding correction $\sigma_{\rm rel, 1ph}$
for different nuclear charges and coefficients of its $\Za$ expansion,
for the point nuclear model.
\label{tab:1ph}}
\begin{ruledtabular}
\begin{tabular}{cw{3.13}}
   $Z$ & \multicolumn{1}{c}{$\sigma_{\rm rel, 1ph}/\alpha^2$}
   \\
\hline\\[-5pt]
  2 &        -0.208\,223\,677 \\
  3 &        -0.208\,086\,391 \\
  4 &        -0.207\,893\,786 \\
  6 &        -0.207\,340\,881 \\
  8 &        -0.206\,560\,303 \\
 10 &        -0.205\,545\,466 \\
 14 &        -0.202\,776\,655 \\
 18 &        -0.198\,940\,386 \\
 22 &        -0.193\,905\,594 \\
 28 &        -0.183\,713\,434 \\
 34 &        -0.169\,616\,681 \\
 40 &        -0.150\,469\,629 \\
 46 &        -0.124\,653\,329 \\
 52 &        -0.089\,862\,182 \\
 \hline
 $d_0$ &  \multicolumn{1}{c}{$-\nicefrac{5}{24}$}\\
 $d_2$ &  \multicolumn{1}{c}{$    0.514\,442\,6$} \\
 $d_4$ &  \multicolumn{1}{c}{$    1.693\,21$} \\
 $d_6$ &  \multicolumn{1}{c}{$    2.469$} \\
% $d_8$ &  \multicolumn{1}{c}{$    3.55$} \\
    \end{tabular}
\end{ruledtabular}
\end{table}

\subsection{QED effects}

In the independent-particle approximation, the QED
contribution for the $(1s)^2$ state
is twice the $1s$ QED correction calculated
for hydrogen-like ions to all orders in $\Za$
in Refs.~\cite{yerokhin:11:prl,yerokhin:12:shield}.
The analytical result for the
leading $\Za$-expansion term was obtained in Refs.~\cite{yerokhin:11:prl,yerokhin:12:shield}
and later corrected in Ref.~\cite{wehrli:21}. Separating out the
leading-order result, we represent
the one-electron QED contribution for the $(1s)^2$ state of helium-like ions as
\begin{align}
\sigma_{\rm QED, 1el} = &\, 2\,\alpha^2\, (\Za)^3
\Big[ -\frac{16}{9\pi}\ln (\Za)
 \nonumber \\ &
-1.896\,642\,389
+  G_{\rm QED}(\Za)
 \Big]\,,
\end{align}
where $G_{\rm QED}(\Za) \approx 2.182\,(\Za) + O((\Za)^2)$ is the remainder function
containing one-electron contributions of higher orders in $\Za$.
The logarithmic term in the above formula corresponds
to the $1/Z^0$ term of the expansion of $\sigma^{(5)}_{\rm log}$,
whereas the other terms have not been included
in the NRQED treatment of Sec.~\ref{sec:1}.
The remainder function $G_{\rm QED}(\Za)$ was calculated to all orders in
$\Za$ in Refs.~\cite{yerokhin:11:prl,yerokhin:12:shield}.
In the present work, we use the data obtained in those works and ascribe
to it the relative uncertainty of $\pm 0.3\,(\Za)^2$, to
account for the uncalculated diagrams with magnetic-loop vacuum-polarization.

So, the higher-order QED contribution beyond those included in Sec.~\ref{sec:1} is
\begin{align}
\sigma_{\rm QED, 1el}^{\rm h.o.} = &\, 2\,\alpha^2\, (\Za)^3
\Big[ -1.896\,642\,389
+  G_{\rm QED}(\Za)
 \Big]\,.
\end{align}

In order to estimate the effects of the screening of the
one-electron QED correction by the second electron,
we introduce the screening factor  $\zeta_{\rm scr}$ basing on
known results for the logarithmic QED contribution $\sigma^{(5)}_{\rm log}$.
Specifically, we define
\begin{align}
\zeta_{\rm scr} = -\frac{\sigma^{(5)}_{\rm log}/[Z^3\ln(\Za)]-c_0}{c_0}\,,
\end{align}
where $c_0 = -\nicefrac{32}{9\pi}$ is the leading $1/Z$-expansion coefficient,
see Table~\ref{tab:NRQED}.
Using this screening factor, we estimate the QED screening contribution as
\begin{align} \label{eq:scr}
\sigma^{\rm h.o.}_{\rm QED, 1ph+} \approx -\zeta_{\rm scr}\,\sigma^{\rm h.o.}_{\rm QED, 1el} \pm 30\%\,.
\end{align}
The above estimate of uncertainty is supported by the complete NRQED calculation of
$\sigma^{(5)}$ for helium \cite{wehrli:21}.

\subsection{Nuclear magnetization distribution}

Within the independent-particle approximation, the nuclear magnetization distribution
correction for the $(1s)^2$ state
is twice the $1s$ hydrogen-like contribution, derived to the leading order in
$\Za$ in Ref.~\cite{pachucki:23:shield}.
Formulas presented in Ref.~\cite{pachucki:23:shield} include both the finite nuclear
size (fns) and the nuclear magnetization distribution (Bohr-Weisskopf, BW) effects.
Removing the fns part, we get
\begin{align}
\sigma_{\rm BW, 1el} = &\, 2\,\Big( -\frac{2\alpha(\Za)^3}{3}\Big) \Big[ m^2r_M^2 + 4\Za\,m \big(r_Z-\lbr r \rbr\big) \Big]\,,
\label{eq:21}
\end{align}
where $r_M$ and $r_Z$ are the magnetic and the Zemach radius, respectively
and $\lbr r \rbr$ is the mean nuclear charge radius.
Within the Gaussian model for the nuclear charge distribution
$\rho_C(r) = \rho_0 \, \exp[-3\,r^2/(2r_C^2)]$ (and similar to the magnetization distribution),
we obtain
\begin{align} \label{eq:21a}
r_Z = \sqrt{\frac{8}{3\pi}}\sqrt{r_C^2 + r_M^2}\,,\ \ \ \lbr r\rbr = \sqrt{\frac{8}{3\pi}}\,r_C\,,
\end{align}
where $r_C = \lbr r^2 \rbr^{1/2}$ is the root-mean-square nuclear charge radius.

For light nuclei with $Z = 3$ and $4$,
we use Eq.~(\ref{eq:21}) with the experimental values of the
Zemach radii, $r_Z(^7\mathrm{Li}) = 3.33$~fm
\cite{pachucki:23:liplushfs} and $r_Z(^9\mathrm{Be}) = 4.04$~fm
\cite{puchalski:14:beplus}.
For heavier nuclei, the Zemach radius is not readily available
from experiment.
For some nuclei,
the $1s$ shielding BW correction was
calculated in Ref.~\cite{yerokhin:12:shield}
within the effective single-particle model of the nuclear magnetization
distribution. However, this model is not universal and is applicable
for some selected nuclei only.
In the present work we will use Eq.~(\ref{eq:21}) with the
magnetic radius expressed in terms of the charge radius by $
r_M = \sqrt{3}\,r_C$ for nuclear charges $Z < 80$.
We found that with this choice of the magnetic radius, Eq.~(\ref{eq:21})
qualitatively reproduces results of the single-particle model calculations of
Ref.~\cite{yerokhin:12:shield}.
We estimate the uncertainty of this approximation of the
one-electron BW correction to be 50\%, which
can be compared to the 30\% uncertainty estimate of the single-particle
model results in Ref.~\cite{yerokhin:12:shield}.
For $Z > 80$, Eq.~(\ref{eq:21}) is no longer adequate. We
thus apply the single-particle nuclear model as described in
Ref.~\cite{yerokhin:12:shield} to compute the BW correction
for several high-$Z$ ions, specifically, with $Z = 80$, $83$, and $91$.

The effects of the electron-electron interaction on the one-electron
BW corrections are estimated analogously to Eq.~(\ref{eq:scr}).

\section{Results and discussion}
\label{sec:3}

In this work we performed numerical calculations of the nuclear shielding correction for the ground state
of helium-like ions for a wide range of nuclear charges. The nuclear parameters were taken
from Ref.~\cite{stone:05} (magnetic moments), Ref.~\cite{angeli:13} (charge radii), and
Ref.~\cite{wang:21} (masses). Individual shielding contributions for
selected ions are presented in Table~\ref{tab:breakdown}. We observe that for the lightest ions,
the dominant theoretical uncertainty comes from the QED screening effect. This uncertainty
can be improved further by a calculation of the nonlogarithmic $\alpha^5$ QED correction, as
accomplished for helium in Ref.~\cite{wehrli:21}. For heavier ions,
the largest theoretical uncertainty
comes from the extended distribution of the nuclear magnetic moment (the BW effect). This uncertainty
can in principle be improved by dedicated calculations of the BW correction for specific nuclei
with a microscopic nuclear model \cite{valuev:phd}. 
An even better way is to use experimental values of
the effective Zemach radius $\tilde{r}_Z$ \cite{pachucki:23:liplushfs} obtained from
the hyperfine-splitting measurements. 
One can then use  $\tilde{r}_Z$ instead
of $r_Z$ in Eq.~(\ref{eq:21}) and compute $r_M$ from Eq.~(\ref{eq:21a}).
An additional benefit is that this automatically accounts for
some higher-order nuclear effects included into $\tilde{r}_Z$. 

Table~\ref{tab:tot} lists our theoretical predictions of the nuclear shielding constant
for helium-like ions.
We do not present results for neutral helium since more complete calculations are available
in this case
\cite{wehrli:21, pachucki:23:shield}. The absolute accuracy of theoretical predictions for the shielding
constant $\sigma$ varies from $2\times 10^{-10}$ for $Z = 3$ to $1\times 10^{-4}$ for $Z = 91$.
This accuracy demonstrates the precision possible for
determination of nuclear magnetic moments from helium-like ions.

Summarizing, we performed calculations of the nuclear magnetic shielding for helium-like ions
in the ground state. By combining two complementary approaches, we obtained results
for a wide region of nuclear charges. Our calculations confirmed the presence of
a rare anti-screening effect for the relativistic shielding correction. They also demonstrated
the importance of inclusion of the negative-energy part of the Dirac spectrum
in calculations of the nuclear shielding, especially for low-$Z$ ions.
In future, the developed approach can be extended to calculations of nuclear shielding
in Li-like ions, which are of immediate experimental interest
\cite{pachucki:10,pachucki:23:limag,dickopf:priv}.

\begin{table*}
\caption{
Individual contributions to the shielding constant in He-like ions,
units are $10^{-6}$.
\label{tab:breakdown}}
\begin{ruledtabular}
\begin{tabular}{lw{3.8}w{3.8}w{3.7}w{3.6}w{4.5}w{4.4}}
   Contribution & \multicolumn{1}{c}{$^7\mathrm{Li}^{+}$}
                    & \multicolumn{1}{c}{$^{9}\mathrm{Be}^{2+}$}
                        & \multicolumn{1}{c}{$^{17}\mathrm{O}^{6+}$}
                            & \multicolumn{1}{c}{$^{43}\mathrm{Ca}^{18+}$}
                                & \multicolumn{1}{c}{$^{73}\mathrm{Ge}^{30+}$}
                                    & \multicolumn{1}{c}{$^{129}\mathrm{Xe}^{52+}$}
   \\
\hline\\[-5pt]
$\sigma^{(2)}$            &   95.423\,74        &  130.918\,47       &   272.9155       &  698.924      &1124.94      &1906.0 \\
$\sigma^{(2,1)}$          &   -0.013\,36        &   -0.020\,13       &    -0.0229       &   -0.029      &  -0.04      &  -0.0 \\
$\sigma^{(4)}$            &    0.158\,52        &    0.359\,23       &     2.7198       &   41.378      & 168.48      & 806.5 \\
$\sigma^{(5)}_{\rm log}$  &    0.001\,79        &    0.004\,24       &     0.0306       &    0.346      &   1.09      &   3.4 \\
$\sigma^{\rm h.o.}_{\rm rel,1el}$
                          & 0.000\,04        &    0.000\,23       &     0.0102       &    1.150      &  12.92      & 204.2
\\
$\sigma^{\rm h.o.}_{\rm rel,1ph+}$
                          & 0.000\,02\,(2)   &    0.000\,06\,(5)  &     0.0010\,(4)  &    0.041\,(6) &   0.29\,(3)    &   2.7\,(2)
\\
$\sigma^{\rm h.o.}_{\rm QED}$
                          &   -0.001\,53\,(16)  &   -0.003\,90\,(29) &    -0.0341\,(12) &   -0.544\,(8) &  -2.23\,(4) & -11.4\,(5)
\\
$\sigma^{\rm h.o.}_{\rm BW }$
                          &   -0.000\,02\,(1)   &   -0.000\,12\,(6)  &    -0.0029\,(14) &   -0.148\,(74)&  -1.13\,(57)& -10.8\,(54)
\\
$\sigma                $  &   95.569\,20\,(16)  &  131.258\,09\,(30) &   275.6172\,(19) &  741.119\,(75)& 1304.30\,(57)&2900.5\,(54)
       \\
    \end{tabular}
\end{ruledtabular}
\end{table*}

\begin{table*}
\caption{
 Nuclear magnetic shielding constant $\sigma$ for the ground state of helium-like ions
 with the nuclear charge number $Z$ and mass number $A$.
 \label{tab:tot}}
\begin{ruledtabular}
\begin{tabular}{llw{3.12}llw{3.10}llw{3.8}}
\multicolumn{1}{c}{$Z$}  &
       \multicolumn{1}{c}{$A$}  &
                 \multicolumn{1}{c}{$\sigma \times 10^3$}
&
\multicolumn{1}{c}{$Z$}  &
       \multicolumn{1}{c}{$A$}  &
                 \multicolumn{1}{c}{$\sigma \times 10^3$}
&
\multicolumn{1}{c}{$Z$}  &
       \multicolumn{1}{c}{$A$}  &
                 \multicolumn{1}{c}{$\sigma \times 10^3$}
\\
\hline\\            %
  3  &    7  &  0.095\,569\,20\,(16)   &  30  &   67  &  1.200\,56\,(43)   &  56  &  137  &  3.106\,4\,(64)  \\
  4  &    9  &  0.131\,258\,09\,(30)   &  31  &   69  &  1.251\,86\,(49)   &  57  &  139  &  3.214\,8\,(69)  \\
  5  &   11  &  0.167\,087\,88\,(50)   &  32  &   73  &  1.304\,30\,(57)   &  58  &  139  &  3.327\,0\,(74)  \\
  6  &   13  &  0.203\,069\,60\,(79)   &  33  &   75  &  1.358\,04\,(65)   &  59  &  141  &  3.443\,3\,(79)  \\
  7  &   14  &  0.239\,245\,6\,(12)    &  34  &   77  &  1.413\,05\,(74)   &  60  &  143  &  3.563\,5\,(85)  \\
  8  &   17  &  0.275\,617\,2\,(19)    &  35  &   79  &  1.469\,43\,(83)   &  62  &  149  &  3.816\,9\,(99)  \\
  9  &   19  &  0.312\,259\,7\,(29)    &  36  &   83  &  1.527\,20\,(94)   &  63  &  151  &  3.951\,(11)  \\
 10  &   21  &  0.349\,167\,0\,(44)    &  37  &   85  &  1.586\,5\,(11)   &  64  &  155  &  4.089\,(12)  \\
 11  &   23  &  0.386\,411\,0\,(63)    &  38  &   87  &  1.647\,3\,(12)   &  65  &  159  &  4.235\,(12)  \\
 12  &   25  &  0.423\,972\,7\,(88)    &  39  &   89  &  1.709\,8\,(13)   &  66  &  161  &  4.382\,(13)  \\
 13  &   27  &  0.461\,945\,(12)       &  40  &   91  &  1.774\,0\,(15)   &  67  &  165  &  4.539\,(14)  \\
 14  &   29  &  0.500\,306\,(16)       &  41  &   93  &  1.839\,9\,(16)   &  68  &  167  &  4.699\,(15)  \\
 15  &   31  &  0.539\,126\,(22)       &  42  &   95  &  1.907\,6\,(18)   &  69  &  169  &  4.869\,(16)  \\
 17  &   35  &  0.618\,236\,(38)       &  43  &   99  &  1.977\,3\,(20)   &  70  &  171  &  5.043\,(17)  \\
 18  &   39  &  0.658\,585\,(48)       &  44  &  101  &  2.049\,0\,(22)   &  71  &  175  &  5.222\,(18)  \\
 19  &   39  &  0.699\,566\,(60)       &  45  &  103  &  2.122\,9\,(25)   &  72  &  177  &  5.414\,(19)  \\
 20  &   43  &  0.741\,119\,(75)       &  46  &  105  &  2.199\,0\,(27)   &  73  &  181  &  5.611\,(20)  \\
 21  &   45  &  0.783\,380\,(92)       &  47  &  107  &  2.277\,4\,(30)   &  74  &  183  &  5.817\,(22)  \\
 22  &   47  &  0.826\,32\,(11)        &  48  &  111  &  2.358\,2\,(32)   &  75  &  185  &  6.033\,(23)  \\
 23  &   51  &  0.870\,04\,(13)        &  49  &  113  &  2.441\,6\,(35)   &  76  &  187  &  6.255\,(24)  \\
 24  &   53  &  0.914\,52\,(16)        &  50  &  119  &  2.527\,6\,(39)   &  77  &  191  &  6.489\,(26)  \\
 25  &   55  &  0.959\,85\,(19)        &  51  &  121  &  2.616\,3\,(42)   &  78  &  195  &  6.731\,(27)  \\
 26  &   57  &  1.006\,07\,(23)        &  52  &  125  &  2.707\,9\,(46)   &  79  &  197  &  6.985\,(29)  \\
 27  &   59  &  1.053\,18\,(27)        &  53  &  127  &  2.802\,6\,(50)   &  80  &  199  &  7.07\,(12)  \\
 28  &   61  &  1.101\,27\,(31)        &  54  &  129  &  2.900\,5\,(54)   &  83  &  209  &  8.102\,(42)  \\
 29  &   63  &  1.150\,39\,(37)        &  55  &  133  &  3.001\,7\,(59)   &  91  &  231  &  10.92\,(13)  \\
    \end{tabular}
\end{ruledtabular}
\end{table*}

%\bibliographystyle{c:/-a-/papers/bibtex/phaip30}
%\bibliography{c:/-a-/papers/bibtex/hfst}

\end{document}